\documentclass[11pt]{article}
\usepackage{epsfig,amssymb,latexsym,amsmath,cite}

\newcommand{\be}{\begin{equation}}
\newcommand{\en}{\end{equation}}
\begin{document}

\thispagestyle{empty}

\baselineskip=15pt

\vspace*{0.5cm}
\begin{center}
{\Large \bf Coherent states for Hamiltonians generated by\\[1ex]
supersymmetry}
\end{center}

\vskip1cm

\begin{center}
{\bf David J. Fern\'andez}$^{\dag}$, {\bf V\'eronique
Hussin}$^{\ddag}$, {\bf Oscar Rosas-Ortiz}$^{\dag}$

\vskip1ex

{\footnotesize \dag\ Departamento de F\'{\i}sica, Cinvestav, AP
14-740, 07000 M\'exico~D~F, Mexico\\[0.5ex]
\ddag\ D\'epartement de Math\'ematiques et Centre de Recherches
Math\'ematiques, Universit\'e de Montr\'eal, C.P. 6128, Succ.
Centre-Ville, Montr\'eal (Qu\'ebec), H3C 3J7, Canada}
\end{center}

\vskip1cm
\begin{center}
\begin{minipage}{12cm}
{\footnotesize {\bf Abstract} Coherent states are derived for
one-dimensional systems generated by supersymmetry  from an
initial Hamiltonian with a purely discrete spectrum for which the
levels depend analytically on their subindex. It is shown that the
algebra of the initial system is inherited by its SUSY partners in
the subspace associated to the isospectral part or the spectrum.
The technique is applied to the harmonic oscillator, infinite well
and trigonometric P\"oschl-Teller potentials. }
\end{minipage}
\end{center}


\section{Introduction}

The great interest in the study of coherent states (CS) stems from the
beautiful properties that the so-called standard ones have, which are a
natural consequence of the huge symmetry supplied by the Heisenberg-Weyl
algebra ruling the harmonic oscillator. Indeed, these characteristics
suggested Glauber to model light by means of standard coherent states
\cite{gl06}, which was a breakthrough in the development of quantum
optics, one of most successful branches of the physics of 20-th century
(see, e.g. \cite{ks85,pe86,zfg90,aag00,do02,ni06}).

Among the several definitions available in the literature for
general systems, algebraically the most important ones are those
which define the CS either as eigenstates of annihilation operators
or as resulting of a `displacement' operator acting onto certain
extremal state. In order to derive the CS following the first
definition, one has to identify the appropriate algebra ruling the
system Hamiltonian, and to find then the annihilation and creation
operators suitable to perform the construction. Since typically the
resulting algebra is not linear, it is usual to call them nonlinear
coherent states
\cite{sp95,dv96,mmsz97,rr99,qu99,si00,sbjps00,rt04,hn05}.

For Hamiltonians $H_k$ generated by supersymmetric quantum mechanics
(SUSY QM)
\cite{ais93,bs97,fe97,bgbm99,ast01,as03,lp03,mr04,in04,gt04,su05,ff05},
the CS analysis has been focussed mainly on the SUSY partners of the
harmonic oscillator \cite{fhn94,kk96,bs96,as97,sbl98,fh99} (see
however \cite{fa93,sa98}). The key ingredient in the approach
introduced in \cite{fhn94,fh99} is to construct a {\it natural} pair
of annihilation and creation operators of $H_k$ simply as products
of intertwining and standard annihilation and creation operators. An
important conclusion of these works was that the natural algebra
ruling the SUSY partner Hamiltonians of the oscillator is a
polynomial deformation of the Heisenberg-Weyl algebra.

For the SUSY partners of a general initial potential, an appropriate
algebraic treatment of the corresponding Hamiltonian $H_0$, ensuring
a right identification of the annihilation and creation operators,
had not been realized. However, for a set of one-dimensional
Hamiltonians with a purely discrete spectrum for which the levels
depend analytically on their index, an {\it intrinsic} algebra has
been identified recently, allowing to calculate in a simple way the
corresponding CS \cite{dh02}. Let us notice that this intrinsic
algebra is in general nonlinear. One of the results of the present
paper is to show that such algebraic structures can be linearized:
one can associate to those systems the Heisenberg-Weyl algebra.
Consequently, an additional set of CS will be constructed, their
explicit expressions containing small variations from the standard
harmonic oscillator CS.

It is remarkable that \cite{dh02} draws as well the attention to the
main subject of this paper, namely, the CS analysis for the SUSY
partners of arbitrary potentials in the spirit of \cite{fhn94,fh99}.
In this context several novel results will be found, e.g., we will
show that the nonlinear and linear algebras of $H_0$ are inherited
by its SUSY partners $H_k$ in the subspace associated to the
isospectral part of the spectrum. In addition, we will find a {\it
natural} algebra for which the generators are products of
annihilation and creation operators of $H_0$ times the intertwiners
of $H_0$ and $H_k$, thus generalizing the previous results for the
harmonic oscillator \cite{fhn94,fh99}. The corresponding CS will be
built up for the several algebras of $H_k$ we are going to study.
Our procedure will be illustrated with the harmonic oscillator,
infinite well and trigonometric P\"oschl-Teller potentials. The
results for the SUSY partners of the infinite well and trigonometric
P\"oschl-Teller potentials, as far as we know, are new.

Let us observe that for specific potentials, like trigonometric
P\"oschl-Teller, Morse and others, there are alternative methods of
construction of CS which employ the symmetry of the differential
equations related to $H_0$ (see e.g. \cite{spb04}). However, to
implement the SUSY transformations departing from such treatments
seems involved, as compared with the technique which will be
presented in this paper (based on \cite{dh02}).

In the next section the initial Hamiltonian we deal with as well as
its related algebras will be studied. The CS analysis for the
several algebras of $H_0$ is the subject of section 3. A brief
overview of SUSY QM as a technique for generating solvable
potentials from a given initial one will be presented in section 4.
In section 5, a pair of nonlinear algebras ruling the SUSY partner
potentials will be discussed, while in section 6 we will explore the
corresponding linear structure. The CS construction for the several
algebras associated to the SUSY partner potentials will be performed
in section 7. In section 8 our general results will be illustrated
with some examples. Finally, in section 9 we close the paper with
our conclusions.

\section{Algebraic structures of the initial Hamiltonian $H_0$}

Let us suppose that the initial system is described by a Hermitian
Schr\"odinger Hamiltonian
\begin{equation}
H_0 = -\frac12 \frac{d^2}{dx^2} + V_0(x),
\end{equation}
whose eigenvectors and eigenvalues satisfy:
\begin{eqnarray}
&& H_0 \vert\psi_n\rangle = E_n \vert\psi_n\rangle,
\label{eqeigenh0} \qquad E_0<E_1<E_2< \dots
\end{eqnarray}
We assume that there is an analytic dependence, defined by a certain
function $E(n)$, of the eigenvalues with the index labeling them,
namely,
\begin{equation}
E_n\equiv E(n),
\end{equation}
and the eigenvectors satisfy the standard orthonormality and completeness
relationships
\begin{equation}
\langle \psi_m\vert \psi_n\rangle = \delta_{mn}, \qquad
\sum_{m=0}^{\infty} \vert\psi_m\rangle\langle\psi_m\vert = 1,
\end{equation}
where the symbol $1$ in any operator expression of this paper
represents the identity operator. There will be
different forms of $E(n)$ according to the system under study,
for instance, for the harmonic oscillator
it will be a linear function of $n$, for an infinite square well it
will be quadratic, etc. This function defines an {\it intrinsic}
algebra which will be next discussed.

\subsection{Intrinsic nonlinear algebra of $H_0$}

Let us define a pair of annihilation and creation operators
$a_0^\pm$ by
\begin{eqnarray}
& a_0^- \vert\psi_n\rangle = r_{\cal I}(n) \vert\psi_{n-1}\rangle,
\quad a_0^+ \vert\psi_n\rangle =  \bar r_{\cal I}(n+1)
\vert\psi_{n+1}\rangle, \label{appsi} \\
& r_{\cal I}(n) = e^{i\alpha(E_{n}-E_{n-1})} \ \sqrt{E_{n}-E_0},
\quad \alpha \in {\mathbb R}, \label{rin}
\end{eqnarray}
such that their product becomes:
\begin{equation}
a_0^+ a_0^- = H_0 - E_0. \label{apam}
\end{equation}
The number operator $N_0$ is now introduced with the properties:
\begin{eqnarray}
&& N_0 \vert\psi_{n}\rangle = n\vert\psi_{n}\rangle, \qquad
[N_0,a_0^\pm] = \pm a_0^\pm. \label{commna}
\end{eqnarray}
As a consequence, two equations which will be widely used along
this work are obtained:
\begin{eqnarray}
&& a_0^\pm g(N_0) = g(N_0\mp 1) a_0^\pm, \label{comaog}
\end{eqnarray}
$g(x)$ being a real arbitrary non-singular function for $x\in
{\mathbb Z}^+$. Combining
Eqs.~(\ref{eqeigenh0},\ref{appsi}-\ref{commna}), it turns out that
the {\it intrinsic algebra} of the system is characterized by the
relationships:
\begin{eqnarray}
&& H_0 = E(N_0), \quad a_0^+ a_0^- = E(N_0) - E_0, \quad a_0^- a_0^+ =
E(N_0+1) - E_0 , \label{n0} \\
&& [a_0^-,a_0^+] = E(N_0+1) - E(N_0) \equiv f(N_0) , \label{n0p1}  \\
&& [H_0,a_0^\pm] = \pm a_0^\pm f\left(N_0 - 1/2 \pm 1/2\right) = \pm
f\left(N_0 - 1/2 \mp 1/2\right) a_0^\pm. \label{coma0h0}
\end{eqnarray}
We will see below that this is not the only algebra of $H_0$ which
can be defined.

Let us notice that we can express $a_0^\pm$ in the form
\begin{eqnarray}
&& a_0^- = r_{\cal I}(N_0 + 1)\sum_{m=0}^{\infty} \vert\psi_m\rangle
\langle\psi_{m+1} \vert , \quad a_0^+ = \bar r_{\cal I}(N_0)
\sum_{m=0}^{\infty} \vert\psi_{m+1}\rangle \langle\psi_{m} \vert ,
\label{apop0}
\end{eqnarray}
where each term in both summations is a Hubbard operator
\cite{hu64,fo89,cph01}. Hence, throughout this paper we will call
these decompositions {\it Hubbard representations}.

\subsection{Linear algebra of $H_0$}

The {\it intrinsic} algebra (\ref{commna},\ref{n0}-\ref{coma0h0})
admits a linearizing procedure, i.e., one can build up new
annihilation and creation operators satisfying the standard
oscillator algebra \cite{fhn94,fh99}. Let us construct them in the
form:
\begin{eqnarray}
&& a_{0_{\mathcal L}}^- = b(N_0) \, a_0^- = a_0^- \, b(N_0 - 1),
\quad a_{0_{\mathcal L}}^+ = a_0^+ \, b(N_0) = b(N_0 - 1) \, a_0^+,
\label{aplineal}
\end{eqnarray}
$b(x)$ being a real non-singular function for $x\in{\mathbb Z}^+$ to
be determined. Suppose that the action of $a_{0_{\mathcal L}}^\pm$
onto the eigenvectors of $H_0$, up to the same phase factors as in
(\ref{appsi}-\ref{rin}), is equal to the oscillator one, namely:
\begin{eqnarray}
&& a_{0_{\mathcal L}}^- \vert\psi_n\rangle = r_{\cal L}(n)
\vert\psi_{n-1}\rangle, \quad a_{0_{\mathcal L}}^+
\vert\psi_n\rangle = \bar r_{\cal L}(n+1) \vert\psi_{n+1}\rangle,
\label{aplinealpsi} \\
&& r_{\cal L}(n) = e^{i\alpha f(n-1)} \, \sqrt{n}. \label{rln}
\end{eqnarray}
On the other hand, the expressions for $a_{0_{\mathcal L}}^\pm$
given in (\ref{aplineal}) and the use of (\ref{appsi}) lead to:
\begin{eqnarray}
&& a_{0_{\mathcal L}}^- \vert\psi_n\rangle = b(n-1) r_{\cal I}(n)
\vert\psi_{n-1}\rangle,  \quad  a_{0_{\mathcal L}}^+
\vert\psi_n\rangle = b(n) \bar r_{\cal I}(n+1)
\vert\psi_{n+1}\rangle \label{aplinealpsi1}.
\end{eqnarray}
By comparing (\ref{aplinealpsi}) with (\ref{aplinealpsi1}) we get:
\begin{equation}
b(n) = \frac{\bar r_{\cal L}(n+1)}{\bar r_{\cal I}(n+1)} =
\frac{r_{\cal L}(n+1)}{r_{\cal I}(n+1)} =
\sqrt{\frac{n+1}{E(n+1)-E_0}}. \label{blineal}
\end{equation}
Making use of (\ref{apop0}-\ref{aplineal},\ref{blineal}), the
Hubbard representation of $a_{0_{\mathcal L}}^\pm$ is obtained:
\begin{eqnarray}
&& a_{0_{\mathcal L}}^- = r_{\cal L}(N_0+1) \sum_{m=0}^{\infty}
\vert \psi_m\rangle \langle \psi_{m+1} \vert, \quad a_{0_{\mathcal
L}}^+ = \bar r_{\cal L}(N_0) \sum_{m=0}^{\infty} \vert
\psi_{m+1}\rangle \langle \psi_{m} \vert, \label{aplinealfinal}
\end{eqnarray}
which, up to the exponential factors of $r_{\cal L}$, is equal to
the oscillator one. Let us notice that, as a consequence of
(\ref{comaog}), we get $a_{0_{\mathcal L}}^\pm g(N_0) = g(N_0\mp 1)
a_{0_{\mathcal L}}^\pm$. Thus, the set $\{N_0, \ a_{0_{\mathcal
L}}^-, \ a_{0_{\mathcal L}}^+\}$ satisfies the oscillator algebra:
\begin{eqnarray}
&& \hskip-1cm [N_0 , a_{0_{\mathcal L}}^\pm] = \pm a_{0_{\mathcal
L}}^\pm, \quad a_{0_{\mathcal L}}^+ a_{0_{\mathcal L}}^- = N_0,
\quad a_{0_{\mathcal L}}^- a_{0_{\mathcal L}}^+ = N_0+1, \quad
[a_{0_{\mathcal L}}^-, a_{0_{\mathcal L}}^+] = 1. \label{comamlapl}
\end{eqnarray}
However, the commutator of $H_0$ with $a^\pm_{0_{\mathcal L}}$
remains the same as for $a_0^\pm$ (see Eq.~(\ref{coma0h0})).

\subsection{General deformation of the intrinsic algebra of $H_0$}

Since it will be used later, it is worth to mention that the
previous linearization is a particular case of a general deformation
of the intrinsic algebra defined by
Eqs.~(\ref{commna},\ref{n0}-\ref{coma0h0}) for $N_0, \ a_0^-, \
a_0^+$. In this procedure, new annihilation and creation operators
$a^- = \beta(N_0) a_0^-$, $a^+ =  a_0^+ \beta(N_0)$, are constructed
such that:
\begin{eqnarray}
&& [N_0 , a^\pm] = \pm a^\pm , \qquad a^+ a^- = \widetilde E(N_0),
\qquad
a^- a^+ = \widetilde E(N_0+1), \\
&& [a^-, a^+] =  \widetilde E(N_0+1) - \widetilde E(N_0) =
\widetilde f(N_0),
\end{eqnarray}
where $\widetilde E(N_0)$ and $\widetilde E(N_0+1)$ are positive
definite operators and $\beta(x)$ is a real non-singular function
for $x\in{\mathbb Z}^+$ to be adjusted according to the chosen
$\widetilde E(N_0)$. It is clear that different choices of
$\widetilde E(N_0)$ lead to different deformations:
\begin{equation}
\hskip-1cm \widetilde E(N_0) = \beta^2(N_0 - 1)[E(N_0)-E_0] \ \
\Rightarrow \ \
\beta(N_0) = \sqrt{\frac{\widetilde E(N_0 + 1)}{E(N_0 + 1) - E_0}}.
\label{gendef}
\end{equation}
In particular, in the previous section we were interested in a
deformation simplifying maximally the original algebra. It can be
here recovered by the choice $\widetilde E(N_0) = N_0$, and by using
(\ref{aplineal},\ref{blineal},\ref{gendef}), it turns out that
$\beta(x) = b(x)$, $a^\pm = a_{0_{\mathcal L}}^\pm$, $\widetilde
f(N_0) = 1$.

\section{Coherent states of $H_0$}

Once some algebras ruling our system have been identified, let us
look for the associated CS. We will derive them as eigenstates of
the several annihilation operators defined previously.

\subsection{Intrinsic nonlinear coherent states of $H_0$}

In the first place, let us analyze the CS $\vert z,\alpha\rangle_0$
which are eigenstates of the annihilation operator of the intrinsic
algebra:
\begin{equation}
a_0^-\vert z,\alpha\rangle_0 = z\vert z,\alpha\rangle_0, \quad
z\in{\mathbb C} . \label{csdef}
\end{equation}
By expanding $\vert z,\alpha\rangle_0$ in the basis of eigenstates
of $H_0$ and following the standard procedure to determine the
expansion coefficients, it turns out that:
\begin{eqnarray}
&& \vert z,\alpha\rangle_0 = \left(\sum_{m=0}^{\infty} \frac{\vert
z\vert^{2m}}{\rho_m}\right)^{-\frac12} \sum_{m =
0}^{\infty}e^{-i\alpha (E_m - E_0)}
\frac{z^m}{\sqrt{\rho_m}}\vert\psi_m\rangle, \label{nlcs0} \\
&& \rho_m =
\begin{cases} 1 & ${\rm if} \ m=0,$
\cr (E_m - E_0)\dots(E_1 - E_0) & ${\rm if} \ $m > 0.
\end{cases}
\label{csdenominator}
\end{eqnarray}

It is important to seek now if the intrinsic nonlinear CS
(\ref{nlcs0}) form a complete set, i.e., if they satisfy
\begin{equation}
\int \vert z,\alpha\rangle_0 \ {}_0\langle z,\alpha \, \vert d\mu(z)
= 1.
\label{completeness1}
\end{equation}
Let us express the positive definite measure $d\mu(z)$ in the form:
\begin{equation}
d\mu(z) = \frac{1}{\pi} \left(\sum_{m=0}^{\infty}\frac{\vert
z\vert^{2m}}{\rho_m}\right) \rho(\vert z\vert^2) \, d^2 z ,
\label{measure}
\end{equation}
$\rho(y)$ being a function to be determined. Working in polar
coordinates and making the change of variable $y=\vert z\vert ^2$,
it is straightforward to show that $\rho(y)$ must satisfy:
\begin{equation}
\int_0^\infty y^m \rho(y) \, dy = \rho_m, \quad m = 0,1,\dots
\label{pm}
\end{equation}
The moment problem (\ref{pm}), in which we look for a positive
definite function $\rho(y)$ with the given $m$-th order moments
$\rho_m$, often arises in the literature when a completeness
relationship of kind (\ref{completeness1}) is to be proven
\cite{bg71,fhn94,fh99,ps99,sp00}. The generic answer is nowadays
known: $\rho(y)$ is the inverse Mellin transform of $\rho_m$
\cite{fh99}. However, for each particular system this calculation
has to be performed explicitly, which is not always easy (see e.g.
\cite{fhn94}).

The expression (\ref{completeness1}) guarantees that any state of
the system can be expanded in terms of CS. In particular, this can
be done for an arbitrary CS $\vert z',\alpha\rangle_0$:
\begin{equation}
\vert z',\alpha\rangle_0 = \int \vert z,\alpha\rangle_0 \ {}_0\langle z,\alpha
\vert z',\alpha\rangle_0 \, d\mu(z), \label{eczpz}
\end{equation}
where the reproducing kernel ${}_0\langle z,\alpha \vert
z',\alpha\rangle_0$ is expressed as:
\begin{equation}
{}_0\langle z,\alpha \vert z',\alpha\rangle_0 = \left(\sum_{m=0}^{\infty}
\frac{\vert z \vert^{2m}}{\rho_m}\right)^{-\frac12} \left(\sum_{m=0}^{\infty}
\frac{\vert z'\vert^{2m}}{\rho_m}\right)^{-\frac12} \left(\sum_{m=0}^{\infty}
\frac{(\bar zz')^m}{\rho_m}\right). \label{reproducing}
\end{equation}
Let us notice that the eigenvalue $z=0$ of $a_0^-$ is
non-degenerated since:
\begin{equation}
\vert z=0,\alpha\rangle_0 = \vert\psi_0\rangle.
\end{equation}
Another important property of the intrinsic nonlinear CS $\vert
z,\alpha\rangle_0$, which is due to the phase choice of
Eqs.~(\ref{appsi}-\ref{rin}), is that they evolve coherently in
time:
\begin{equation}
U_0(t) \vert z,\alpha\rangle_0 = e^{-itE_0}\vert z,\alpha+t\rangle_0,
\end{equation}
$U_0(t)=\exp(-itH_0)$ being the evolution operator associated to
$H_0$.

\subsection{Linear coherent states of $H_0$}

Let us study the CS which are eigenstates of the linear annihilation
operator of $H_0$:
\begin{equation}
a_{0_{\mathcal L}}^- \vert z,\alpha\rangle_{0_{\mathcal L}} = z
\vert z,\alpha\rangle_{0_{\mathcal L}}, \quad z\in{\mathbb C} .
\label{eigenaml}
\end{equation}
Hence:
\begin{equation}
\vert z,\alpha\rangle_{0_{\mathcal L}} = e^{-\frac{\vert z
\vert^2}2} \sum_{m = 0}^{\infty} e^{-i\alpha(E_m -
E_{0})}\frac{z^m}{\sqrt{m!}} \, \vert\psi_m\rangle.
\label{eclineales}
\end{equation}
Up to the phases involving $\alpha$, they have the form of the
standard harmonic oscillator CS.

Contrasting with the difficulty to find a positive definite measure
ensuring the completeness of the non-linear CS (\ref{nlcs0}), now
the problem is already solved:
\begin{equation}
\frac{1}{\pi} \int \vert z,\alpha\rangle_{0_{\mathcal L}} \
{}_{0_{\mathcal L}}\langle z, \alpha \vert \, d^2 z =1,
\label{completeinil}
\end{equation}
i.e., the measure is the standard one, $d^2z/\pi$. Thus, an
arbitrary linear CS $\vert z',\alpha\rangle_{0_{\mathcal L}}$ admits
a non-trivial decomposition in terms of $\vert
z,\alpha\rangle_{0_{\mathcal L}}$:
\begin{equation}
\vert z',\alpha\rangle_{0_{\mathcal L}} = \frac{1}{\pi}\int \vert
z,\alpha\rangle_{0_{\mathcal L}} \ {}_{0_{\mathcal L}}\langle z,
\alpha \vert z',\alpha\rangle_{0_{\mathcal L}} \, d^2 z,
\end{equation}
where the reproducing kernel is equal to the oscillator one:
\begin{equation}
{}_{0_{\mathcal L}}\langle z, \alpha \vert
z',\alpha\rangle_{0_{\mathcal L}} = \exp\left(-\frac{\vert z
\vert^2}2 + \bar z z' - \frac{\vert z' \vert^2}2 \right).
\label{kerlin}
\end{equation}
The only eigenstate of $H_0$ which is as well a linear CS
(\ref{eclineales}) is again the ground state:
\begin{equation}
\vert z=0,\alpha\rangle_{0_{\mathcal L}} = \vert\psi_0\rangle.
\end{equation}
Since $[a_{0_{\mathcal L}}^-,a_{0_{\mathcal L}}^+]=1$, the linear CS
also result from acting a `displacement' operator onto
$\vert\psi_0\rangle$:
\begin{equation}
\vert z,\alpha\rangle_{0_{\mathcal L}} = D_{\mathcal
L}(z)\vert\psi_0\rangle = \exp(za_{0_{\mathcal L}}^+ - \bar z
a_{0_{\mathcal L}}^-)\vert\psi_0\rangle.
\end{equation}

\section{The SUSY partner Hamiltonians $H_k$}

Let us discuss in the first place some generalities of the SUSY
partner Hamiltonians $H_k$,
\begin{equation}
H_k = - \frac12 \frac{d^2}{dx^2} + V_k(x),
\end{equation}
generated from $H_0$ through a $k$-th order differential
intertwining operator $B_k^+$ \cite{fh99},
\begin{equation}
H_k B_k^+ = B_k^+ H_0 \quad \Leftrightarrow \quad H_0 B_k = B_k H_k.
\label{intrel}
\end{equation}
The potential $V_k(x)$ reads:
\begin{equation}
V_k(x) = V_0(x) - \sum_{i=1}^k \alpha_i'(x,\epsilon_i), \label{nv}
\end{equation}
where, in case that the $k$ factorization energies $\epsilon_i, \ i=
1,\dots,k$ are all different, $\alpha_i(x,\epsilon_i)$ is obtained
from a recursive (B\"acklund) formula:
\begin{equation}
\hskip-1cm \alpha_i(x,\epsilon_i) = -\alpha_{i-1}(x,\epsilon_{i-1}) -
\frac{2(\epsilon_i -
\epsilon_{i-1})}{\alpha_{i-1}(x,\epsilon_{i})-\alpha_{i-1}(x,\epsilon_{i-1})},
\quad  i=2,\dots k, \label{backlund}
\end{equation}
and $\alpha_1(x,\epsilon_i)$ are solutions of the following Riccati
equation:
\begin{equation}
\alpha_1'(x,\epsilon_i) + \alpha_1^2(x,\epsilon_i) = 2 [V_0(x) - \epsilon_i], \quad i=1,\dots,k.
\end{equation}
This is equivalent to the initial stationary Schr\"odinger equation
for the factorization energies $\epsilon_i$, as can be seen from the
change $\alpha_1(x,\epsilon_i)= u_i'(x)/u_i(x)$:
\begin{eqnarray}
& - \frac12 u_i'' + V_0(x)u_i = \epsilon_i u_i. \label{sch}
\end{eqnarray}
In terms of the transformation functions $u_i$, the new potential in
(\ref{nv}) becomes:
\begin{equation}
V_k(x) = V_0(x) - \{\ln[W(u_1,\dots,u_k)]\}'', \label{nvu}
\end{equation}
$W(u_1,\dots,u_k)$ being the Wronskian of the involved solutions of
(\ref{sch}). It is worth to notice that, in order to obtain
nontrivial results when two (or more) factorization energies
coincide, the confluent limit of the previous formulae has to be
used \cite{mnr00,fs03}. It is important also to write down the
relevant factorizations for the SUSY QM of $k$-th order:
\begin{eqnarray}
&& B_k^+ B_k = \prod\limits_{i=1}^k (H_k-\epsilon_i), \qquad B_k
B_k^+  = \prod\limits_{i=1}^k (H_0-\epsilon_i).
\end{eqnarray}

Let us suppose now that, as a result of the $k$-th order
intertwining technique, $s$ of the states annihilated by $B_k$ are
as well physical eigenstates of $H_k$ associated to the eigenvalues
$\epsilon_i$. By convenience, they will be specially denoted by
$\vert \theta_{\epsilon_i}\rangle, \ B_k\vert
\theta_{\epsilon_i}\rangle = 0$, $H_k \vert
\theta_{\epsilon_i}\rangle = \epsilon_i \vert
\theta_{\epsilon_i}\rangle$, $i=1,\dots,s$, $s\leq k$. However, we
assume that the procedure creates just $q$ additional levels with
respect to ${\rm Sp}(H_0)$, but without deleting any of the original
levels of $H_0$, i.e.,
\begin{equation}
{\rm Sp}(H_k) = \{ \epsilon_1, \dots, \epsilon_q, E_0, E_1, \dots
\}, \quad q\leq s.
\end{equation}
This means that $p\equiv s-q$ factorization energies $\epsilon_{q +
j}$ coincide with $p$ energy levels $E_{m_j}$ of $H_0$, i.e.,
$\epsilon_{q+j}= E_{m_j}, \ j=1,\dots,p, \ m_j<m_{j+1}$, and thus
$B_k^+\vert\psi_{m_j}\rangle = 0$. The eigenstates $\vert\theta_n
\rangle$ of $H_k$ which are associated to the remaining energies
$E_n, \ n\neq m_j$, are obtained from the initial ones $\vert
\psi_n\rangle$ and vice versa through the intertwining operators
$B_k^+$ and $B_k$, namely:
\begin{eqnarray}
\vert\theta_n\rangle =  \frac{B_k^+\vert
\psi_n\rangle}{\sqrt{\prod\limits_{i=1}^k(E_n-\epsilon_i)}}, \qquad
 \vert \psi_n\rangle  =  \frac{B_k\vert
\theta_n\rangle}{\sqrt{\prod\limits_{i=1}^k(E_n-\epsilon_i)}}.
\label{psi}
\end{eqnarray}
It is convenient to extend now the definition of $\vert \theta_n
\rangle$ for $n = m_j$ in the way:
\begin{eqnarray}
\vert \theta_{m_j} \rangle \equiv \vert
\theta_{\epsilon_{q+j}}\rangle, \ j=1,\dots,p.
\end{eqnarray}
Summarizing all this information, the eigenstates $\vert \theta_{\epsilon_i}
\rangle, \ \vert \theta_n \rangle$ of $H_k$ obey:
\begin{eqnarray}
&& H_k \vert \theta_n \rangle = E_n \vert \theta_n \rangle, \quad
H_k \vert \theta_{\epsilon_i}\rangle = \epsilon_i \vert
\theta_{\epsilon_i}\rangle,  \\
&& \langle \theta_{\epsilon_i} \vert \theta_n \rangle = 0, \quad
\langle \theta_m \vert  \theta_n \rangle = \delta_{mn}, \quad
\langle \theta_{\epsilon_i} \vert\theta_{\epsilon_j}\rangle =
\delta_{ij},  \\
&& \sum_{l=1}^{s} \vert \theta_{\epsilon_l}\rangle \langle
\theta_{\epsilon_l} \vert + \widetilde{\sum_{m}} \vert \theta_m
\rangle\langle \theta_m\vert = \sum_{l=1}^{q} \vert
\theta_{\epsilon_l}\rangle \langle \theta_{\epsilon_l} \vert +
\sum_{m = 0}^{\infty} \vert \theta_m \rangle\langle \theta_m \vert
=1 ,
\end{eqnarray}
where $n,m=0,1,\dots, \ i,j=1,\dots,q$, $\widetilde\sum_{m}$ is the
sum over $m = 0,1,\dots$ except by $m_j, \ j = 1,\dots,p$, and the
identity operator has been expanded in two alternative ways which
will be useful later. Since the positions of the new levels
$\epsilon_i, \ i=1,\dots,q$ are arbitrary, one might think that some
algebraic properties of $H_0$ are inherited by $H_k$ on the subspace
spanned by the $\vert \theta_n \rangle, \ n=0,1,\dots$ Keeping this
in mind, let us analyze some interesting algebras of the SUSY
partner Hamiltonians $H_k$.

\section{Nonlinear algebras of $H_k$}

We define first a {\it number operator} $N_k$ by its action onto
the eigenstates of $H_k$:
\begin{eqnarray}
&& N_k \vert \theta_n \rangle = n \vert \theta_n \rangle, \quad N_k
\vert \theta_{\epsilon_i}\rangle = 0, \quad n=0,1,\dots \quad
i=1,\dots,q .
\end{eqnarray}
Notice that this definition is more natural than a previous one,
introduced as the ``generalized number operator'' for the SUSY
partners of the oscillator (compare with Eq.~(3.4) of \cite{fh99}).

Let us study next two pairs of annihilation and creation operators
of $H_k$ (and $N_k$) as well as their corresponding nonlinear
algebras.

\subsection{Natural algebra of $H_k$}

Here we will obtain annihilation and creation operators of $H_k$
following a 3-steps construction previously introduced for the SUSY
partner Hamiltonians of the harmonic oscillator
\cite{fhn94,fh99,mi84}. Thus, starting from the {\it intrinsic}
operators $a_0^\pm$ of $H_0$  and the intertwining ones $B_k, \
B_k^+$ of (\ref{intrel}), a pair of {\it natural} annihilation and
creation operators $a^\pm_{k_{\mathcal N}}$ of $H_k$ is built up:
\begin{eqnarray}
&& a^\pm_{k_{\mathcal N}} = B_k^+ a_0^\pm B_k. \label{apnatural}
\end{eqnarray}
Since $B_k\vert \theta_{\epsilon_i}\rangle =0, i=1,\dots,s$, one can
find the action of $a^\pm_{k_{\mathcal N}}$ onto the basis of
eigenvectors of $H_k$ (and $N_k$) by using (\ref{appsi},\ref{psi}):
\begin{eqnarray}
& a^\pm_{k_{\mathcal N}} \vert \theta_{\epsilon_i}\rangle = 0 , \quad
i=1,\dots,q, \label{akntei}\\
& a^-_{k_{\mathcal N}} \vert \theta_n \rangle = r_{\cal N}(n) \
\vert \theta_{n-1} \rangle, \quad a^+_{k_{\mathcal N}} \vert
\theta_n \rangle = \bar r_{\cal N}(n + 1) \ \vert \theta_{n+1}
\rangle, \quad n=0,1,\dots \label{akntn} \\
& r_{\cal N}(n) = \left\{\prod\limits_{i=1}^k [E(n) -
\epsilon_i][E(n-1)-\epsilon_i] \right\}^\frac12 r_{\cal I}(n).
\label{rnnrin}
\end{eqnarray}
Notice that $r_{\cal N}(m_j)=0, j=1,\dots,p$, which is consistent
with $B_k \vert \theta_{m_j}\rangle = a^-_{k_{\mathcal N}} \vert
\theta_{m_j}\rangle = 0$. From these expressions one can find the
Hubbard representation for $a^\pm_{k_{\mathcal N}}$:
\begin{eqnarray}
a^-_{k_{\mathcal N}} & = & r_{\cal N}(N_k+1) \sum_{m=0}^{\infty}
\vert\theta_m \rangle\langle\theta_{m+1} \vert,  \quad
a^+_{k_{\mathcal N}} =  \bar r_{\cal N}(N_k) \sum_{m=0}^{\infty}
\vert\theta_{m+1} \rangle\langle\theta_{m} \vert. \label{aknp}
\end{eqnarray}
Making use of $ a^\pm_{k_{\mathcal N}} g(N_k) = g(N_k \mp 1)
a^\pm_{k_{\mathcal N}}$ for an arbitrary regular function $g(x),
x\in {\mathbb Z}^+$, one can show that:
\begin{eqnarray}
\hskip-0.5cm [a^-_{k_{\mathcal N}}, a^+_{k_{\mathcal N}}] & = &
\left[\bar r_{\cal N}(N_k+1) r_{\cal N}(N_k+1) - \bar r_{\cal
N}(N_k) r_{\cal N}(N_k)\right] \sum_{m=0}^{\infty} \vert\theta_{m}
\rangle\langle\theta_{m} \vert . \label{aknmaknp}
\end{eqnarray}

\subsection{Intrinsic algebra of $H_k$}

It is interesting to observe that simpler annihilation and creation
operators for $H_k$ can be constructed, proceeding by analogy with
(\ref{apop0}). Thus, we define the {\it intrinsic} annihilation and
creation operators $a_k^\pm$ of $H_k$ as follows:
\begin{eqnarray}
&& a_k^- = r_{\cal I}(N_k+1) \sum_{m=0}^{\infty}
\vert\theta_m\rangle \langle\theta_{m+1} \vert , \quad a_k^+ = \bar
r_{\cal I}(N_k) \sum_{m=0}^{\infty} \vert\theta_{m+1}\rangle
\langle\theta_{m} \vert , \label{apopk}
\end{eqnarray}
where $r_{\cal I}(n)$ is given in (\ref{rin}). It can be checked
that $a_k^\pm \vert \theta_{\epsilon_i} \rangle = 0, \ i=1,\dots,q$,
and:
\begin{eqnarray}
&&  a_k^- \vert \theta_n \rangle = r_{\cal I}(n) \vert \theta_{n-1}
\rangle , \quad a_k^+ \vert \theta_n \rangle = \bar r_{\cal I}(n +
1)\vert \theta_{n+1} \rangle,  \label{akptheta} \\
&& a_k^+ a_k^- \vert \theta_n \rangle = (E_n - E_0) \vert \theta_n
\rangle , \quad a_k^- a_k^+ \vert \theta_n \rangle = (E_{n+1} - E_0)
\vert \theta_n \rangle .
\end{eqnarray}
Thus, the commutator between $a_k^\pm$ is similar to that for the
intrinsic algebra of $H_0$ on the subspace spanned by $\{\vert
\theta_n \rangle, \ n=0,1,\dots \}$:
\begin{eqnarray}
&[a_k^-,a_k^+] = f(N_k) \sum\limits_{m=0}^\infty \vert\theta_{m}
\rangle\langle\theta_{m} \vert .
\end{eqnarray}

We would like to seek next if there is any connection between the
initial and final number operators $N_0$ and $N_k$. After some
simple manipulations, it can be shown that:
\begin{eqnarray}
&& \hskip-2cm N_k =  C_k^+ N_0 \, C_k + \sum_{j=1}^{p} m_j \vert
\theta_{m_j}\rangle\langle \theta_{m_j}\vert \ \Leftrightarrow \ N_k
\widetilde{\sum_{m}} \vert
\theta_{m}\rangle\langle \theta_{m}\vert = C_k^+ N_0 \, C_k, \\
C_k & = &
\frac{1}{\sqrt{\prod\limits_{i=1}^k [E(N_0)-\epsilon_i]}} \, B_k, \qquad
C_k^+ = \frac{1}{\sqrt{\prod\limits_{i=1}^k[E(N_k)-\epsilon_i]}} \, B_k^+ ,
\label{ckp}
\end{eqnarray}
$C_k$, $C_k^+$ being {\it modified intertwining operators} inverse
to each other when acting on the eigenstates of the isospectral part
which are not used as seeds in the SUSY procedure, i.e.,
\begin{eqnarray}
&& C_k \vert \theta_n\rangle = \vert \psi_n\rangle ,  \quad C_k^+
\vert \psi_n\rangle = \vert \theta_n\rangle ,  \quad {\mathbb Z}^+
\ni n\neq m_j, \ j=1,\dots,p,\label{ckpsi}
\end{eqnarray}
but in general they are not invertible in the full Hilbert space
${\mathcal L}^2({\mathbb R})$ since $ C_k\vert
\theta_{\epsilon_i}\rangle = C_k\vert \theta_{m_j}\rangle =C_k^+
\vert \psi_{m_j}\rangle = 0, i=1,\dots,q, \ j=1,\dots,p$. From these
expressions one can check that
\begin{eqnarray}
&& a_k^\pm \widetilde{\sum_{m}} \vert \theta_{m}\rangle\langle
\theta_{m}\vert = C_k^+ a_0^\pm \, C_k. \label{akp}
\end{eqnarray}
By using Eqs.~(\ref{ckpsi}-\ref{akp}) one recovers (\ref{akptheta}).
Moreover, it turns out that
\begin{eqnarray}
a_k^+ a_k^- = [E(N_k) - E_0] = [H_k - E_0] \sum_{m=0}^\infty
\vert\theta_{m} \rangle\langle\theta_{m} \vert .
\end{eqnarray}

The RHS of the expressions (\ref{akp}) for the intrinsic operators
$a_k^\pm$ consist of a 3-steps action, similar to the natural ones
$a_{k_{\mathcal N}}^\pm$ of (\ref{apnatural}). The difference is
that the new intertwiners $C_k, \ C_k^+$ transform the states $\vert
\theta_n\rangle \leftrightarrow \vert \psi_n\rangle, \ {\mathbb Z}^+
\ni n\neq m_j, j=1,\dots,p,$ without changing the norm (compare
(\ref{ckpsi}) with (\ref{psi})). This explains why the {\it
intrinsic} algebra generated by $\{N_k, a_k^-,a_k^+\}$ is simpler
than the {\it natural} one obtained from $\{N_k, a_{k_{\mathcal
N}}^-,a_{k_{\mathcal N}}^+\}$. In addition, the intrinsic algebra is
a deformation of the natural one and vice versa (remember section
2.3). Indeed, by comparing (\ref{aknp}) with (\ref{apopk}) one can
show that:
\begin{eqnarray}
&& \hskip-2.3cm a_{k_{\mathcal N}}^- = \frac{r_{\mathcal N}(N_k +
1)}{r_{\mathcal I}(N_k + 1)} \, a_k^-, \quad a_{k_{\mathcal N}}^+ =
\frac{r_{\mathcal N}(N_k)}{r_{\mathcal I}(N_k)} \, a_k^+, \quad
a_{k_{\mathcal N}}^+ a_{k_{\mathcal N}}^- = [E(N_k) - E_0]
\left[\frac{r_{\mathcal N}(N_k)}{r_{\mathcal I}(N_k)} \right]^2
\label{ndihk}
\end{eqnarray}
We will see next another deformation of the intrinsic
algebra generated by $\{N_k, a_k^-,a_k^+\}$.

\section{Linear algebra of $H_k$}

Let us introduce now a new pair of annihilation and creation
operators for $H_k$, such that their action onto the
$\vert\theta_n\rangle$'s is similar to the oscillator one (see
(\ref{aplinealpsi}-\ref{rln})):
\begin{eqnarray}
&& a_{k_{\mathcal L}}^- \vert\theta_n\rangle = r_{\cal L}(n)
\vert\theta_{n-1}\rangle, \nonumber \quad  a_{k_{\mathcal L}}^+
\vert\theta_n\rangle = \bar r_{\cal L}(n+1)
\vert\theta_{n+1}\rangle, \\ && a_{k_{\mathcal L}}^\pm
\vert\theta_{\epsilon_i}\rangle = 0, \quad i=1,\dots,q. \nonumber
\label{apklinealpsi}
\end{eqnarray}
In the Hubbard representation we have:
\begin{eqnarray}
&& a_{k_{\mathcal L}}^- = r_{\cal L}(N_k + 1)
\sum_{m=0}^{\infty}\vert \theta_m\rangle \langle \theta_{m+1} \vert,
\quad a_{k_{\mathcal L}}^+ = \bar r_{\cal L}(N_k)
\sum_{m=0}^{\infty} \vert \theta_{m+1}\rangle \langle \theta_{m}
\vert. \label{aklp}
\end{eqnarray}
It is simple to show that:
\begin{eqnarray}
&& [N_k, a_{k_{\mathcal L}}^\pm] =  \pm a_{k_{\mathcal L}}^\pm ,
\qquad [a_{k_{\mathcal L}}^-, a_{k_{\mathcal L}}^+] =
\sum_{m=0}^\infty \vert \theta_m\rangle\langle \theta_m \vert.
\end{eqnarray}
One can also find that:
\begin{eqnarray}
&& a_{k_{\mathcal L}}^\pm \widetilde{\sum_{m}} \vert
\theta_{m}\rangle\langle \theta_{m}\vert =  C_k^+ a_{0_{\mathcal
L}}^\pm C_k.
\end{eqnarray}

By comparing (\ref{aklp}) with (\ref{apopk}), it is seen that the
linear annihilation and creation operators $a_{k_{\mathcal L}}^\pm$
are deformations of the intrinsic ones $a_{k}^\pm$ to get a simpler
algebra, namely:
\begin{eqnarray}
&& a_{k_{\mathcal L}}^- = \frac{r_{\mathcal L}(N_k + 1)}{r_{\mathcal
I}(N_k + 1)} \, a_{k}^- , \quad a_{k_{\mathcal L}}^+ =
\frac{r_{\mathcal L}(N_k)}{r_{\mathcal I}(N_k)} \, a_{k}^+, \quad
a_{k_{\mathcal L}}^+ a_{k_{\mathcal L}}^- = N_k. \label{ldihk}
\end{eqnarray}

\section{Coherent states of $H_k$}

Let us construct three sets (in general non-equivalent) of CS as
eigenstates of $a_{k_{\mathcal N}}^-, \ a_k^-, \ a_{k_{\mathcal
L}}^-$. According to the algebra involved, they will be called
natural, intrinsic and linear CS respectively. It will be seen that
some differences with respect to the CS of $H_0$ arise.

\subsection{Natural nonlinear coherent states of $H_k$}

We build up first the {\it natural nonlinear coherent states} $\vert
z,\alpha\rangle_{k_{\mathcal N}}$ which are eigenstates of
$a^-_{k_{\mathcal N}}$. Their expansion in terms of
eigenstates of $H_k$ read:
\begin{equation}
\vert z,\alpha\rangle_{k_{\mathcal N}} = \sum_{i=1}^q
\gamma_{\epsilon_i} \vert\theta_{\epsilon_i}\rangle +
\sum_{m=0}^\infty \gamma_m \vert\theta_m\rangle.
\end{equation}
From the CS definition and making use of (\ref{akntei}-\ref{akntn}),
we get $\gamma_{\epsilon_i}=0, \ i=1,\dots,q$, and
\begin{eqnarray}
&& r_{\cal N}(m)\gamma_m = z \gamma_{m-1}, \quad m=1,2,\dots
\label{rrdn}
\end{eqnarray}
According to our SUSY treatment, $\epsilon_s = E_{m_{p}}$ is the
largest eigenvalue of $H_k$, of the part isospectral to $H_0$, for
which $B_k \vert\theta_{m_{p}}\rangle = a^\pm_{k_{\mathcal
N}}\vert\theta_{m_{p}}\rangle = 0$. Moreover, since $B_k^+
\vert\psi_{m_{p}}\rangle = 0$ it turns out that $a^-_{k_{\mathcal
N}}\vert\theta_{m_{p}+1}\rangle = 0$, i.e., $r_{\cal N}(m_{p} + 1) =
0$, and by using (\ref{rrdn}) this implies that $\gamma_{m_{p}} =
0$. By iterating down this equation we arrive at $\gamma_m = 0, \
m=0,\dots,m_{p}$. Eq.~(\ref{rrdn}) can be used again to express
$\gamma_{m + m_{p} + 1}, m>0$, in terms of $\gamma_{m_{p}+1}$:
\begin{equation}
\hskip-0.8cm \gamma_{m + m_{p} + 1} = \frac{z^m}{r_{\cal N}(m +
m_{p} + 1)r_{\cal N}(m + m_{p})\dots r_{\cal N}(m_{p}+2)} \,
\gamma_{m_{p}+1}, \quad m
> 0.
\end{equation}
By using the normalization condition and asking for
$\gamma_{m_{p}+1} \in {\mathbb R}^+$, we finally obtain:
\begin{equation}
\vert z,\alpha\rangle_{k_{\mathcal N}} = \left(\sum_{m = 0}^{\infty}
\frac{\vert z\vert^{2m}}{\widetilde\rho_m}\right)^{-\frac12} \sum_{m
= 0}^{\infty} e^{-i\alpha (E_{m + m_{p} + 1} -
E_{m_{p}+1})}\frac{z^{m}}{\sqrt{\widetilde\rho_m}}\vert\theta_{m +
m_{p} + 1}\rangle , \label{csakn}
\end{equation}
where $\widetilde\rho_0 = 1$ and, for $m  >  0$,
\begin{eqnarray} && \hskip-2cm \widetilde\rho_m  =
\frac{\rho_{m + m_{p} + 1}}{\rho_{m_{p}+1}} \prod\limits_{i=1}^k
(E_{m + m_{p} + 1}  - \epsilon_i) (E_{m + m_{p}}  -
\epsilon_i)^2 \dots (E_{m_{p}+2}  - \epsilon_i)^2(E_{m_{p}+1}  -
\epsilon_i), \label{csakndenominator}
\end{eqnarray}
with $\rho_m$ given by (\ref{csdenominator}).

An important difference of $\vert z,\alpha\rangle_{k_{\mathcal N}}$
with respect to the sets of CS of $H_0$ is that the completeness
relationship now has to include the eigenstates of $H_k$ which are
missing in the expansion (\ref{csakn}), i.e.,
\begin{equation}
\sum_{i=1}^q \vert \theta_{\epsilon_i}\rangle \langle
\theta_{\epsilon_i} \vert + \sum_{m = 0}^{m_{p}} \vert
\theta_m\rangle \langle \theta_m \vert + \int \vert
z,\alpha\rangle_{k_{\mathcal N}} \, {}_{k_{\mathcal N}}\langle
z,\alpha \vert \, d\widetilde\mu(z) = 1.
\end{equation}
A similar procedure as for the CS of $H_0$ leads to:
\begin{equation}
d\widetilde\mu(z) = \frac{1}{\pi} \left(\sum_{m = 0}^{\infty}
\frac{\vert z\vert^{2m}}{\widetilde\rho_m}\right)
\widetilde\rho(\vert z\vert^2) \, d^2z,
\end{equation}
$\widetilde\rho(y)$ satisfying a moment problem more complicated
than the initial one (compare $\rho_m$ of (\ref{csdenominator}) with
$\widetilde\rho_m$ of (\ref{csakndenominator})):
\begin{equation}
\int_0^\infty y^m \widetilde\rho(y) \, dy = \widetilde \rho_m ,
\quad m \geq 0. \label{mpn}
\end{equation}
Another relevant difference is that, since $B_k \vert
\theta_{\epsilon_i}\rangle = a^-_{k_{\mathcal N}} \vert
\theta_{\epsilon_i}\rangle = 0, \ i=1,\dots,q$, $B_k \vert
\theta_{m_{j}}\rangle = a^-_{k_{\mathcal N}} \vert
\theta_{m_{j}}\rangle = 0$, $a^-_{k_{\mathcal N}} \vert
\theta_{m_{j}+1}\rangle = 0, \ j=1,\dots p$, and $a^-_{k_{\mathcal
N}} \vert \theta_{0}\rangle = 0$, then the degeneracy of the
eigenvalue $z=0$ of $a^-_{k_{\mathcal N}}$ can be any integer in the
set $\{s+1, \dots,s+p+1 \}$, depending on the positions of the
levels $E_{m_{j}}, \ j=1,\dots, p$. However, once again by the phase
choice of Eq.~(\ref{rin}), the natural CS $\vert
z,\alpha\rangle_{k_{\mathcal N}}$ of (\ref{csakn}) evolve coherently
in time:
\begin{equation}
U_k(t) \vert z,\alpha\rangle_{k_{\mathcal N}} = e^{-itE_{m_{p}+1}}
\vert z,\alpha + t\rangle_{k_{\mathcal N}},
\end{equation}
$U_k(t) = \exp (-it H_k)$ being the evolution operator associated to
$H_k$. This property also will be valid for the other CS of $H_k$
which will be further derived.

Let us remark that some properties of the natural nonlinear CS of
$H_k$ were studied previously for the SUSY partners of the harmonic
oscillator \cite{fhn94,fh99}. To compare with the case discussed in
\cite{fh99}, let us restrict ourselves to SUSY transformations for
which the seeds are just nonphysical eigenfunctions of $H_0$, i.e.,
take $p=0$ and $q=s\leq k$. Now the only eigenstate of $H_k$ for the
part of the spectrum isospectral to $H_0$ which is annihilated by
$a_{k_{\mathcal N}}^-$ is $\vert\theta_0\rangle$, and thus the CS
expansion (\ref{csakn}) should start from this state. This is
achieved by defining $m_{p=0}=-1$: with this choice and taking the
harmonic oscillator energy levels in the CS of (\ref{csakn}) one
arrives to the CS of Eq.~(5.14) in \cite{fh99}.

\subsection{Intrinsic nonlinear coherent states of $H_k$}

Let us analyze next the intrinsic nonlinear CS $\vert
z,\alpha\rangle_k$ which are eigenstates of $a_k^-$. A similar
procedure as before leads to
\begin{equation}
\vert z,\alpha\rangle_k = \left(\sum_{m=0}^{\infty} \frac{\vert
z\vert^{2m}}{\rho_m}\right)^{-\frac12} \sum_{m = 0}^{\infty}
e^{-i\alpha (E_m -
E_0)}\frac{z^m}{\sqrt{\rho_m}}\vert\theta_m\rangle . \label{csak}
\end{equation}
This expansion is also obtained from the intrinsic nonlinear CS
$\vert z,\alpha\rangle_0$ of $H_0$ and vice versa by the change
$\vert\psi_n\rangle \leftrightarrow \vert\theta_n\rangle$ (compare
Eqs.~(\ref{nlcs0}) and (\ref{csak})). Thus, the completeness
relationship is automatically satisfied,
\begin{equation}
\sum_{i=1}^q \vert \theta_{\epsilon_i}\rangle \langle
\theta_{\epsilon_i} \vert + \int \vert z,\alpha\rangle_k \,
{}_k\langle z,\alpha \vert \, d\mu(z) = 1,
\end{equation}
where $d\mu(z)$ is given by equations (\ref{measure},\ref{pm}). This
is a simplification with respect to the natural nonlinear CS $\vert
z,\alpha\rangle_{k_{\mathcal N}}$ of
(\ref{csakn},\ref{csakndenominator}). After some simple
manipulations we also arrive at
\begin{equation}
\vert z,\alpha\rangle_k = C_k^+ \vert z,\alpha\rangle_0 +
\left(\sum_{m=0}^{\infty} \frac{\vert
z\vert^{2m}}{\rho_m}\right)^{-\frac12} \sum_{j=1}^{p} e^{-i\alpha
(E_{m_j} -
E_0)}\frac{z^{m_j}}{\sqrt{\rho_{m_j}}}\vert\theta_{m_j}\rangle.
\end{equation}
Since $a_k^- \vert \theta_{\epsilon_i} \rangle = 0, \ i=1,\dots,q$
and taking into account that
\begin{equation}
\vert z=0,\alpha\rangle_k = \vert \theta_{0}\rangle,
\end{equation}
it turns out that the eigenvalue $z=0$ of $a_k^-$ is $(q+1)$-th
degenerated.

\subsection{Linear coherent states of $H_k$}

Let us consider the linear CS which are eigenstates of
$a_{k_{\mathcal L}}^-$. Since the algebra of $a_{k_{\mathcal
L}}^\pm$ acting onto ${\rm Span}\{\vert\theta_n\rangle,
n=0,1,\dots\}$ is equal to that of $a_{0_{\mathcal L}}^\pm$ acting
onto ${\rm Span}\{\vert\psi_n\rangle, n=0,1,\dots\}$, it can be
shown that:
\begin{equation}
\vert z,\alpha\rangle_{k_{\mathcal L}} = e^{-\frac{\vert z
\vert^2}2} \sum_{m = 0}^{\infty} e^{-i\alpha(E_m -
E_{0})}\frac{z^m}{\sqrt{m!}} \, \vert\theta_m\rangle.
\label{ecklineales}
\end{equation}
This expression is also obtained from the corresponding one for
$\vert z,\alpha\rangle_{0_{\mathcal L}}$ and vice versa by the
mapping $\vert\psi_m\rangle \leftrightarrow \vert\theta_m\rangle$
(compare (\ref{eclineales}) and (\ref{ecklineales})). Thus, the
completeness relationship is identified in a simple way:
\begin{equation}
\sum_{i=1}^q \vert \theta_{\epsilon_i}\rangle \langle
\theta_{\epsilon_i} \vert + \frac{1}{\pi}\int \vert
z,\alpha\rangle_{k_{\mathcal L}} \, {}_{k_{\mathcal L}}\langle
z,\alpha \vert \, d^2z = 1.
\end{equation}
Moreover, it turns out that:
\begin{equation}
\vert z,\alpha\rangle_{k_{\mathcal L}} = C_k^+\vert
z,\alpha\rangle_{0_{\mathcal L}} + e^{-\frac{\vert z \vert^2}2}
\sum_{j=1}^{p} e^{-i\alpha(E_{m_j} -
E_{0})}\frac{z^{m_j}}{\sqrt{{m_j}!}} \, \vert\theta_{m_j}\rangle.
\end{equation}
The eigenvalue $z=0$ of $a_{k_{\mathcal L}}^-$ is $(q+1)$-th
degenerated, a property discovered for the first time for the SUSY
partners of the harmonic oscillator \cite{fhn94,fh99}. It can also
be found that
\begin{equation}
\vert z,\alpha\rangle_{k_{\mathcal L}} = D_{k_{\mathcal L}}\vert
\theta_{0}\rangle = \exp(za_{k_{\mathcal L}}^+ -  \bar z
a_{k_{\mathcal L}}^-)\vert \theta_{0}\rangle.
\end{equation}

\section{Examples}

We will apply the previous techniques to some examples: the harmonic
oscillator, infinite square well and trigonometric P\"oschl-Teller
potentials. For each system we will use a different kind of SUSY
transformation, depending on how many physical eigenstates
$\vert\theta_{\epsilon_i}\rangle$ of $H_k$ which are annihilated by
$B_k$ have energies different from the ones of $H_0$. Thus, for the
harmonic oscillator we will study the general situation with $q\neq
0, \ p\neq 0$, while for the infinite square well the strictly
isospectral case with $q=0, \ p=s$ will be explored. For the
P\"oschl-Teller potential the $s$ levels $\epsilon_i$ will be
different from the ones of $H_0$ (i.e. for $q=s, \ p=0$).

\subsection{The harmonic oscillator}

Let us consider the harmonic oscillator potential:
\begin{equation}
V_0(x) = \frac{x^2}{2}. \label{ho}
\end{equation}
The normalized eigenfunctions and eigenvalues of $H_0$ are given by:
\begin{eqnarray}
&&\hskip-0.5cm \psi_n(x) = \langle x \vert \psi_n\rangle =
\frac{e^{-\frac{x^2}2} H_n(x)}{\sqrt{\sqrt{\pi}2^nn!}}, \quad E(n)
\equiv E_n = n + \frac12, \quad n=0,1,\dots \label{psin}
\end{eqnarray}
where $H_n(x)$ are the Hermite polynomials. Since $E(n)$ is linear
in $n$, it is simple to show that $f(N_0) = 1$. Thus, after dropping
some unimportant global phases, the {\it intrinsic} algebra reduces
to the Heisenberg-Weyl one, as it was expected. This implies that
the corresponding CS become as well the canonical ones (take $\alpha
= 0$ in the formulae of sections 2.1 and 3.2).

\subsubsection{The SUSY partners $H_k$.}

Let us study the $k$-th order SUSY partners of the harmonic
oscillator. In order to implement the transformation, we look for
the general solution $u(x)$ of the stationary Schr\"odinger equation
(\ref{sch}) with the oscillator potential (\ref{ho}) for an
arbitrary factorization energy $\epsilon$. Up to a constant factor
we obtain:
\begin{equation}
\hskip-1cm u(x) = e^{-\frac{x^2}2} \left[{}_1 F_1 \left(\frac14 -
\frac{\epsilon}2; \frac12; x^2\right) + 2\mu x \frac{\Gamma(\frac34
- \frac{\epsilon}2)}{\Gamma(\frac14 - \frac{\epsilon}2)} \ {}_1 F_1
\left(\frac34 - \frac{\epsilon}2; \frac32; x^2\right)\right],
\label{sgsho}
\end{equation}
where ${}_1 F_1(a;b;y)$ is the confluent hypergeometric function and
$u(x)$ is nodeless for $\epsilon < 1/2, \ \vert \mu \vert < 1$
\cite{fh99}. By using this expression to specify the seed solutions,
the associated Wronskian can be calculated, which automatically
leads to the new potential and the corresponding energy eigenstates.

\subsubsection{Algebraic structures of $H_k$.}

The annihilation and creation operators for the several algebras of
$H_k$, in terms of the intrinsic ones $a_k^\pm$, are given by
Eqs.~(\ref{ndihk},\ref{ldihk}), where:
\begin{eqnarray}
& \frac{r_{\cal N}(n)}{r_{\cal I}(n)} = \left[
\prod\limits_{i=1}^k \left(n - \epsilon_i - \frac12\right)\left(n
- \epsilon_i + \frac12 \right)\right]^{\frac12}, \quad
\frac{r_{\cal L}(n)}{r_{\cal I}(n)} = 1.
\end{eqnarray}
Up to a global phase factor, the intrinsic operators $a_k^\pm$ are
those of (\ref{apopk}) with $r_{\cal I}(n) = \sqrt{n}$, i.e., we
recover the Heisenberg-Weyl algebra onto ${\rm
Span}\{\vert\theta_n\rangle, n=0,1,\dots\}$.

\subsubsection{Coherent states of $H_k$.}

In order to find the natural nonlinear CS of $H_k$, we determine
first the coefficients $\widetilde \rho_m$ of
(\ref{csakn},\ref{csakndenominator}):
\begin{equation}
\widetilde\rho_m  = (m_{p} + 2)_m \prod\limits_{i=1}^k \left(m_{p}-
\epsilon_i+\frac32\right)_m \left(m_{p}-\epsilon_i+\frac52\right)_m,
\quad m\geq 0, \label{mnho}
\end{equation}
with the Pochhammer symbol given by $(b)_m = \Gamma(b+m)/\Gamma(b)$.
Hence we get:
\begin{eqnarray}
&& \hskip-1.5cm \vert z,\alpha\rangle_{k_{\mathcal N}} = \frac{1}{\sqrt{
{}_1F_{2k+1}(1;m_p+2,\dots,m_p-\epsilon_i+\frac32,m_p-\epsilon_i +
\frac52,\dots ;\vert z \vert^2)}} \nonumber \\ &&  \times \sum_{m =
0}^{\infty}  \frac{z^m
\vert\theta_{m+m_p+1}\rangle}{\sqrt{(m_p+2)_m} \prod\limits_{i=1}^k
\sqrt{(m_p-\epsilon_i+\frac32)_m(m_p-\epsilon_i+\frac52)_m}}
  , \label{csaknho}
\end{eqnarray}
where ${}_p F_q$ is a generalized hypergeometric function defined
by:
\begin{eqnarray}
&& {}_p \, F_q(a_1,\dots,a_p;b_1,\cdots,b_q;x) = \sum_{m = 0}^\infty
\frac{(a_1)_m\dots(a_p)_m}{(b_1)_m\cdots(b_q)_m} \frac{x^m}{m!}.
\end{eqnarray}
It is clear that the moment problem (\ref{mpn}) with the
$\widetilde\rho_m$ of (\ref{mnho}) is more involved than the already
solved initial one, and it can be worked out once the factorization
energies $\epsilon_i$ are specified. Indeed, a few solutions for
some SUSY transformations have been derived elsewhere
\cite{fhn94,fh99}.

For the intrinsic nonlinear and linear CS of $H_k$, both expressions
are the same and coincide with the canonical expansion, which arises
from (\ref{eclineales}) for $\alpha = 0$ with the change
$\vert\psi_m\rangle \rightarrow \vert\theta_m\rangle$.

In particular, we illustrate the SUSY partner potential $\widetilde
V_3(x)$ generated from a third-order transformation with $k = 3$, $q
= p = 1$. The seeds $u_1, \ u_2, \ u_3$, correspond to the solution
(\ref{sgsho}) with $\epsilon_1=-3/2$ for $u_1$, the ground state
eigenfunction $\psi_0(x)$ of (\ref{psin}) with $\epsilon_2 = E_0 =
1/2$ for $u_2$, and a generalized eigenfunction of second order
associated to $\epsilon_3 = \epsilon_2$ for $u_3$ such that $(H_0 -
\epsilon_2)u_3 = u_2 \Rightarrow (H_0 - \epsilon_2)^2 u_3 = 0$, its
nontrivial part given by \cite{fs03}:
\begin{equation}
u_3= \frac{e^{-\frac{x^2}2}}{2\pi^\frac14} \left[\pi w_0 {\rm
Erfi}(x) + x^2 \ {}_2F_2\left(1,1;\frac32,2;x^2\right)\right].
\end{equation}
The new potential is obtained from (\ref{nvu}), with the Wronskian
expressed as:
\begin{eqnarray}
\hskip-1.0cm W(u_1,u_2,u_3) & = & \frac{e^{-\frac{3x^2}2}}{
\sqrt{\pi}} \bigg\{ -2x + 4\pi w_0\mu x e^{2x^2} +
\sqrt{\pi}e^{x^2}\bigg[4w_0 - \mu -2 \mu x^2 \nonumber \\
&& \hskip-1.5cm  + \bigg(1+2\sqrt{\pi}(\mu + 2 w_0)x e^{x^2} -
2x^2\bigg){\rm Erf}(x)\bigg] + 2 \pi x e^{2x^2}[{\rm Erf}(x)]^2
\bigg\}
\end{eqnarray}
This Wronskian is nodeless for $\vert\mu\vert<1$ and $\vert
w_0\vert>1/2$. A member of the family of potentials (\ref{nvu}) is
shown in Fig. 1 for $\mu = 0.99$ and $w_0 = 0.51$. The spectrum of
the Hamiltonian $H_3$ is $\{\epsilon_1=-3/2, \ E_n = n + 1/2, \
n=0,1,\dots\}$.

\begin{figure}[ht]
\centering \epsfig{file=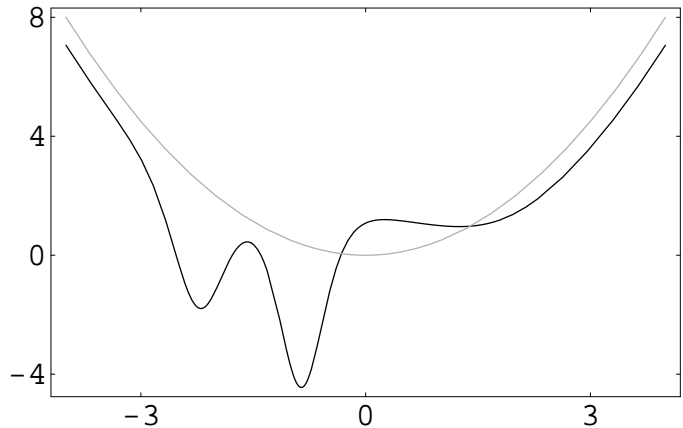, width=10cm} \caption{\small
Third-order SUSY partner potential $V_3(x)$ (black curve) of the
oscillator (gray curve) obtained by composing a confluent
second-order transformation with seed the ground state of $H_0$
($w_0 = 0.51$) and a first-order one with $\epsilon_1=-3/2$ ($\mu =
0.99$). The net result is the `creation' of an energy level at
$\epsilon_1$ for $H_3$.}
\end{figure}

\subsection{The infinite well potential}

In dimensionless units, the infinite well potential we shall study
reads:
\begin{equation}
V_0(x) = \left\{
\begin{array}{cl}
\infty & \mbox{\rm for} \quad x = 0, \pi \\
0 & \mbox{\rm for} \quad  0 < x < \pi.
\end{array}
\right.
\label{viw}
\end{equation}
The eigenfunctions and eigenvalues are well known:
\begin{eqnarray}
&& \hskip-1cm \psi_n(x)  = \sqrt{\frac{2}{\pi}} \,
\sin\left[(n+1)x\right], \quad E_n = E(n) = \frac{(n+1)^2}{2}, \quad
n=0,1,\dots \label{psiniw}
\end{eqnarray}

\subsubsection{Intrinsic algebra of $H_0$.}It is determined by the
operator function
\begin{equation}
E(N_0) = \frac{(N_0 + 1)^2}2 = H_0 ,
\end{equation}
leading thus to the following structure function:
\begin{equation}
f(N_0) = E(N_0 + 1) - E(N_0) = N_0 + \frac32  . \label{fiw}
\end{equation}
The Hubbard representation for the intrinsic operators $a_0^\pm$ is
given by (\ref{apop0}), where now:
\begin{eqnarray}
&& r_{\cal I}(n) = e^{i\alpha \left(n + \frac12\right)} \
\sqrt{\frac{n(n + 2)}{2}}. \label{riniw}
\end{eqnarray}
The operator set $\{ N_0, a_0^-,a_0^+\}$ satisfies then the
commutation relationships:
\begin{eqnarray}
&& [N_0, a_0^\pm] =  \pm a_{0}^\pm , \qquad [a_{0}^-, a_{0}^+] = N_0
+ \frac32 ,
\end{eqnarray}
which, after redefining the number operator as $\widetilde N_0 = N_0
+ \frac32$, reduce to the {\rm su}(1,1) algebra.

\subsubsection{Linear algebra of $H_0$.}

The linear operators $a_{0_{\mathcal L}}^\pm$, expressed as
deformations of the intrinsic ones $a_{0}^\pm$, acquire the form:
\begin{eqnarray}
&& a_{0_{\mathcal L}}^- = \sqrt{\frac{2}{N_0 + 3}} \ a_{0}^- ,
\qquad a_{0_{\mathcal L}}^+ = a_{0}^+ \sqrt{\frac{2}{N_0 + 3}},
\qquad a_{0_{\mathcal L}}^+ a_{0_{\mathcal L}}^- = N_0.
\end{eqnarray}
By construction, their action onto the eigenstates of $H_0$ is the
standard one (up to some phase factors).

\subsubsection{Coherent states of $H_0$.}

The intrinsic nonlinear and linear CS of $H_0$ become:
\begin{eqnarray}
&& \vert z,\alpha \rangle_{0} = \left[ {}_0F_1(3;2\vert
z\vert^2)\right]^{-\frac12} \sum_{m = 0}^\infty e^{-i
\frac{\alpha}2m(m+2)} \sqrt{\frac{2^{m+1}}{m!\, (m+2)!}} \ z^m
\vert\psi_m\rangle ,  \label{csiwi} \\ && \vert z,\alpha
\rangle_{0_{\mathcal L}} = e^{-\frac{\vert z \vert^2}2} \sum_{m =
0}^{\infty} e^{-i \frac{\alpha}2m(m+2)}\frac{z^m}{\sqrt{m!}} \,
\vert\psi_m\rangle . \label{csiwl}
\end{eqnarray}
The completeness of the intrinsic nonlinear CS (\ref{csiwi}) is
ensured since the moment problem (\ref{pm}) with $\rho_m = m! \,
(m+2)!/2^{m+1}$ admits the positive definite solution
\begin{equation}
\rho(y) = 4 y K_2\left(2\sqrt{2y}\right),
\end{equation}
$K_2(y)$ being a modified Bessel function of second kind. Hence, the
measure (\ref{measure}) reads:
\begin{equation}
d\mu(z) = \frac{4 \vert z\vert^2}{\pi} \, {}_0 F_1(3;2\vert
z\vert^2) \,  K_2(2\sqrt{2} \, \vert z\vert) \, d^2 z .
\end{equation}
The reproducing kernel (\ref{reproducing}) acquires the form:
\begin{eqnarray}
&& {}_0\langle z,\alpha \vert z',\alpha\rangle_0 = \left[{}_0
F_1(3;2\vert z\vert^2) \ {}_0 F_1(3;2\vert
z'\vert^2)\right]^{-\frac12} {}_0 F_1(3;2 \bar zz').
\end{eqnarray}

On the other hand, for the linear CS (\ref{csiwl}) directly apply
the formulae of section 3.2, in particular the completeness
relationship (\ref{completeinil}) and the reproducing kernel
(\ref{kerlin}).

\subsubsection{The SUSY partners $H_k$.}

For generating the $k$-th order SUSY partners of the infinite well
potential, we employ isospectral transformations which do not create
new levels. This implies that $q=0, \ p = s \leq k$, and there are
$p$ levels of $H_0$, $\epsilon_j = E_{m_j}= (m_j+1)^2/2, \
j=1,\dots,p$, whose physical eigenstates $\vert\psi_{m_j}\rangle$
are annihilated by $B_k^+$ and will be used as seeds to implement
the procedure.

\subsubsection{Algebraic structures of $H_k$.}

The natural and linear annihilation and creation operators of $H_k$,
in terms of the intrinsic ones $a_k^\pm$, are written in
Eqs.~(\ref{ndihk},\ref{ldihk}), where:
\begin{eqnarray}
& \frac{r_{\cal N}(n)}{r_{\cal I}(n)} = 2^{-k}\prod\limits_{i=1}^k
\sqrt{[n^2 - 2\epsilon_i][(n + 1)^2 - 2\epsilon_i]}, \quad
\frac{r_{\cal L}(n)}{r_{\cal I}(n)} = \sqrt{\frac{2}{n + 2}}.
\end{eqnarray}
The intrinsic operators are given in Eq.~(\ref{apopk}) with $r_{\cal
I}(n)$ given by (\ref{riniw}).

\subsubsection{Coherent states of $H_k$.}

The coefficients $\widetilde \rho_m$ in
(\ref{csakn},\ref{csakndenominator}), required to find the natural
nonlinear CS $\vert z,\alpha\rangle_{k_{\mathcal N}}$, take the
form:
\begin{eqnarray}
&&  \hskip-1.5cm  \widetilde \rho_m =  \frac{(m_p+2)_m
(m_p+4)_m}{2^{m(2k+1)}}\prod\limits_{i=1}^k (m_p -
\sqrt{2\epsilon_i} + 2)_m(m_p - \sqrt{2\epsilon_i} + 3)_m
\nonumber \\ && \hskip3.0cm \times (m_p + \sqrt{2\epsilon_i} +
2)_m(m_p + \sqrt{2\epsilon_i} + 3)_m, \quad m\geq 0. \label{nrho}
\end{eqnarray}
Therefore:

{\scriptsize
\begin{eqnarray}
\vert z,\alpha\rangle_{k_{\mathcal N}} = \mbox{\hskip12cm}
\nonumber\\[2ex]
\frac{1}{\sqrt{{}_1F_{4k+2}(1;m_p \! + \! 2,m_p \! + \! 4,\dots,m_p
\! - \! \sqrt{2\epsilon_i} \! + \! 2,m_p \! - \! \sqrt{2\epsilon_i}
\! + \! 3,m_p \! + \! \sqrt{2\epsilon_i} \! + \! 2,m_p \! + \!
\sqrt{2\epsilon_i} \! + \! 3, \dots ;2^{2k+1}\vert z \vert^2)}}
\nonumber\\[2ex]
\times \! \! \sum_{m = 0}^{\infty} \!\!
\frac{e^{-\frac{i}2\alpha m(m+2m_p+4)}\sqrt{2^{m(2k+1)}} \,
z^m\vert\theta_{m+m_p+1}\rangle}{\sqrt{(m_p \! + \!2)_m(m_p \! + \!
4)_m} \! \prod\limits_{i=1}^k\!\! \sqrt{(m_p \! - \!
\sqrt{2\epsilon_i} \! + \! 2)_m (m_p \! - \! \sqrt{2\epsilon_i} \! +
\! 3)_m (m_p \! + \! \sqrt{2\epsilon_i} \! + \! 2)_m (m_p \! + \!
\sqrt{2\epsilon_i} \! + \! 3)_m}}
\end{eqnarray}
}

\noindent
The moment problem (\ref{mpn}) with the $\widetilde\rho_m$ of
(\ref{nrho}) can be worked out once the factorization energies
$\epsilon_1, \dots, \epsilon_k$ are specified. These quantities
determine as well the degeneracy of the eigenvalue $z=0$ of
$a_{k_{\mathcal N}}$, which can take a value in the set
$\{p+1,\dots,2p+1\}$.

The intrinsic nonlinear and linear CS of $H_k$ are obtained from
(\ref{csiwi}) and (\ref{csiwl}) respectively by the replacement
$\vert\psi_m\rangle \rightarrow \vert\theta_m\rangle$.

For illustrating some isospectral SUSY partners of the infinite
well (\ref{viw}), we employ a confluent second-order
transformation involving one physical eigenfunction of $H_0$,
i.e., we take $k=2, \ \epsilon_1 = \epsilon_2 = E_{m_1} =
(m_1+1)^2/2$ \cite{mnr00,fs03}. We need to evaluate the Wronskian
of two generalized eigenfunctions $u_1, \ u_2$ of $H_0$: $u_1$ is
the standard physical eigenfunction $\psi_{m_1}(x)$ of
(\ref{psiniw}) obeying $(H_0 - \epsilon_1)u_1 = 0$, but $u_2$ is a
second-order generalized eigenfunction such that $(H_0 -
\epsilon_1)u_2 = u_1 \Rightarrow (H_0 - \epsilon_1)^2u_2 = 0$
\cite{fs03}. The expression for $u_2$ is:
\begin{equation}
u_2(x) = - \frac{(\pi w_0 + x)}{\sqrt{2\pi}({m_1}+1)}
\cos[({m_1}+1)x].
\end{equation}
This allows to evaluate the Wronskian  $W(u_1,u_2)$, and then the
new potential,

{\scriptsize
\begin{equation}
V_2(x) = \left\{
\begin{array}{cc}
\infty & \mbox{\rm for} \quad x =0, \pi\\[1ex]
\frac{16({m_1}+1)^2\sin[({m_1}+1)x]\{\sin[({m_1}+1)x]-
({m_1}+1)(\pi w_0 + x)\cos[({m_1}+1)x]\}} {\{\sin[2({m_1}+1)x]-2
({m_1}+1)(\pi w_0 + x)\}^2} & \mbox{\rm for} \quad 0<x<\pi,
\end{array}
\right.
\label{viwk}
\end{equation}}

\noindent which is non-singular for $x\in(0,\pi)$ if $w_0>0$ or
$w_0<-1$. An example of these potentials is shown in Fig. 2 for
$m_1=1$, $w_0=0.1$ (black curve), where it is drawn in gray the
infinite well (\ref{viw}).

\begin{figure}[ht]
\centering \epsfig{file=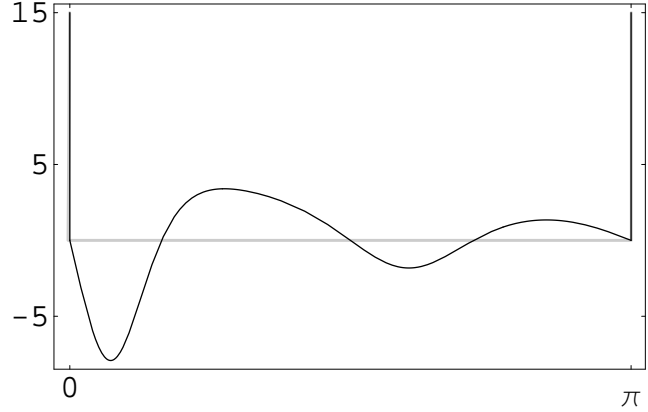, width=10cm} \caption{\small
Second-order SUSY partner potential $V_2(x)$ (black curve)
isospectral to the infinite well (gray line) obtained by a confluent
second-order transformation involving the eigenfunction of the first
excited state of $H_0$ and $w_0=0.1$.}
\end{figure}

\subsection{The trigonometric P\"oschl-Teller potential}

In appropriate units the trigonometric P\"oschl-Teller potential can
be written:
\begin{equation}
V_0(x) = \frac{\nu(\nu-1)}{2\cos^2(x)} , \quad  \nu>1 .
\label{vpt}
\end{equation}
The energy eigenstates $\psi_{n}(x)$ are expressed in terms of
Gegenbauer polynomials $C_n^\nu(y)$ while the eigenvalues are
quadratic in $n$ \cite{ni78,qu99}:
\begin{eqnarray}\label{2.29}
&& {\psi_n}(x)= \left[\frac{n!(n +
\nu)\Gamma(\nu)\Gamma(2\nu)}{\sqrt{\pi}\,\Gamma(\nu +
\frac12)\Gamma(n + 2\nu)}\right]^{1/2} \
\cos^{\nu}(x) \ C_n^\nu(\sin(x)),\nonumber \\
&& E_{n} = E(n) = \frac{(n+\nu)^2}2, \quad n=0,1,2,\ldots
\end{eqnarray}

\subsubsection{Intrinsic algebra of $H_0$.}It is defined by:
\begin{equation}
E(N_0) = \frac{(N_0 + \nu)^2}2 = H_0 ,
\end{equation}
giving place to the following structure function:
\begin{equation}
f(N_0) = E(N_0 + 1) - E(N_0) = N_0 + \nu + \frac12. \label{fnpt}
\end{equation}
The Hubbard representation for the intrinsic operators $a_0^\pm$ is
given again by (\ref{apop0}) with:
\begin{eqnarray}
&&  r_{\cal I}(n) = e^{i\alpha \left(n + \nu - \frac12\right)} \
\sqrt{\frac{n(n + 2\nu)}{2}}. \label{rinpt}
\end{eqnarray}
The operator set $\{ N_0, a_0^-,a_0^+\}$ satisfies the commutation
relationships:
\begin{eqnarray}
&& [N_0, a_0^\pm] =  \pm a_{0}^\pm , \qquad [a_{0}^-, a_{0}^+] =
N_0 + \nu + \frac12 ,
\end{eqnarray}
which, redefining the number operator as $\widetilde N_0 = N_0 + \nu
+ \frac12$, reduce to the ${\rm su}(1,1)$ algebra.

\subsubsection{Linear algebra of $H_0$.}

The linear annihilation and creation operators $a_{0_{\mathcal
L}}^\pm$ can be expressed as deformations of the intrinsic ones
$a_{0}^\pm$:
\begin{eqnarray}
&& \hskip-1cm a_{0_{\mathcal L}}^- = \sqrt{\frac{2}{N_0 + 2\nu +1}}
\ a_{0}^- , \quad a_{0_{\mathcal L}}^+ = a_{0}^+  \sqrt{\frac{2}{N_0
+ 2\nu + 1}}, \quad a_{0_{\mathcal L}}^+ a_{0_{\mathcal L}}^- = N_0.
\end{eqnarray}
Once again, by construction they act on the eigenstates of $H_0$ in
a standard way (up to some phase factors).

\subsubsection{Coherent states of $H_0$.}

The intrinsic nonlinear and linear CS become now:
\begin{eqnarray}
&&\hskip-1cm \vert z,\alpha \rangle_{0} = \left[
{}_0F_1(2\nu+1;2\vert z\vert^2)\right]^{-\frac12} \sum_{m =
0}^\infty e^{-i \frac{\alpha}2m(m+2\nu)}
\sqrt{\frac{2^m}{m!(2\nu+1)_m}} \ z^m
\vert\psi_m\rangle ,  \label{cspti} \\
&& \vert z,\alpha \rangle_{0_{\mathcal L}} = e^{-\frac{\vert z
\vert^2}2} \sum_{m = 0}^{\infty} e^{-i
\frac{\alpha}2m(m+2\nu)}\frac{z^m}{\sqrt{m!}} \, \vert\psi_m\rangle
. \label{csptl}
\end{eqnarray}
The set of intrinsic nonlinear CS (\ref{cspti}) is complete since
the moment problem (\ref{pm}) with
\begin{eqnarray}
&& \rho_m = \frac{m! \, (2\nu+1)_m}{2^m}
\end{eqnarray}
can be simply solved, with a positive definite function $\rho(y)$
given by:
\begin{eqnarray}
&& \rho(y) = \frac{2^{\nu+2}y^\nu}{\Gamma(2\nu+1)}
K_{2\nu}(2\sqrt{2y}).
\end{eqnarray}
Hence, the invariant measure (\ref{measure}) becomes:
\begin{equation}
d\mu(z) = \frac{2^{\nu+2}\vert z\vert^{2\nu}}{\pi\Gamma(2\nu+1)} \,
{}_0 F_1(2\nu+1; 2\vert z\vert^2) K_{2\nu}(2\sqrt{2} \, \vert
z\vert) \, d^2 z .
\end{equation}
The reproducing kernel (\ref{reproducing}) reads:
\begin{eqnarray}
& \hskip-1cm {}_0\langle z,\alpha \vert z',\alpha\rangle_0 \!=\!
\left[{}_0 F_1(2\nu+1;2\vert z\vert^2) \ {}_0 F_1(2\nu+1;2\vert
z'\vert^2)\right]^{-\frac12} {}_0 F_1(2\nu+1;2 \bar zz').
\end{eqnarray}

For the linear CS (\ref{csptl}) of $H_0$ all formulae of section 3.2
become the same, so we skipped them, as we did for the infinite well
potential (\ref{viw}).

\subsubsection{The SUSY partners $H_k$.}

For generating the $k$-th order SUSY partners of the
P\"oschl-Teller potential (\ref{vpt}), we use
transformations involving just seed solutions associated to
non-physical factorization energies $\epsilon_i, \ i=1,\dots,k$, of
$H_0$, $q$ of them becoming physical levels of $H_k$. The
general mathematical eigenfunction $u(x)$ of $H_0$ for arbitrary
$\epsilon$ is given by:
\begin{eqnarray}
&& \hskip-1.5cm u(x) = \cos^{\nu}(x)\bigg[
{}_2F_1\left(\frac{\nu}{2}-\sqrt{\frac{\epsilon}{2}},
\frac{\nu}{2} +
\sqrt{\frac{\epsilon}{2}}; \frac12;\sin^2(x)\right) \nonumber \\
&& \hskip1.0cm  + \mu \ \sin(x) \ {}_2F_1\left(\frac{\nu}{
2}+\sqrt{\frac{\epsilon}{2}} + \frac12,
\frac{\nu}{2}-\sqrt{\frac{\epsilon}{2}}+ \frac12;
\frac32;\sin^2(x)\right)\bigg]. \label{ptgs}
\end{eqnarray}
This expression supplies any seed solution involved in the Wronskian
of the transformation, which leads to the potential $V_k(x)$ as well
as the eigenstates of $H_k$.

\subsubsection{Algebraic structures of $H_k$.}

The annihilation and creation operators for the natural and linear
algebras of $H_k$ are written in Eqs.~(\ref{ndihk},\ref{ldihk})
with:
\begin{eqnarray}
& \hskip-1.5cm \frac{r_{\cal N}(n)}{r_{\cal I}(n)} =
2^{-k}\prod\limits_{i=1}^k \sqrt{[(n + \nu - 1)^2 -
2\epsilon_i][(n + \nu)^2 - 2\epsilon_i]}, \quad \frac{r_{\cal
L}(n)}{r_{\cal I}(n)} = \sqrt{\frac{2}{n + 2\nu}}.
\end{eqnarray}
The intrinsic operators are given in Eq.~(\ref{apopk}) with $r_{\cal
I}(n)$ given by (\ref{rinpt}).

\subsubsection{Coherent states of $H_k$.}

The coefficients $\widetilde \rho_m$ of
(\ref{csakn},\ref{csakndenominator}) required to find the natural
nonlinear CS of $H_k$ are now: {\small\begin{eqnarray} &&
\hskip-1.5cm \widetilde \rho_m  = \frac{m! (2\nu +
1)_m}{2^{m(2k+1)}}\prod_{i=1}^k (\nu  - \sqrt{2\epsilon_i})_m(\nu  -
\sqrt{2\epsilon_i} + 1)_m (\nu  + \sqrt{2\epsilon_i})_m(\nu  +
\sqrt{2\epsilon_i} + 1)_m, \label{nrhopt}
\end{eqnarray}}
where $m\geq 0$. Therefore:
{\small\begin{eqnarray} && \hskip-2.3cm
\vert z,\alpha\rangle_{k_{\mathcal N}} =
\frac{1}{\sqrt{ {}_0F_{4k+1}(2\nu \! + \! 1, \dots,\nu \! - \!
\sqrt{2\epsilon_i},\nu \! - \! \sqrt{2\epsilon_i} \! + \! 1,\nu \! +
\!
\sqrt{2\epsilon_i},\nu \! + \! \sqrt{2\epsilon_i} \! + \! 1, \dots ;2^{2k+1}\vert z \vert^2)}} \nonumber \\
&& \hskip-1.2cm \times \! \! \sum_{m = 0}^{\infty}\!\!
\frac{e^{-\frac{i}2\alpha m(m + 2\nu)}\sqrt{2^{m(2k+1)}} \,
z^m\vert\theta_{m}\rangle}{\sqrt{m!(2\nu \! + \! 1)_m} \!
\prod\limits_{i=1}^k \!\!\sqrt{\left(\nu \! - \!
\sqrt{2\epsilon_i}\right)_m \! \left(\nu \! - \! \sqrt{2\epsilon_i}
\! + \! 1\right)_m \! \left(\nu \! + \! \sqrt{2\epsilon_i}\right)_m
\! \left(\nu \! + \! \sqrt{2\epsilon_i} \! + \! 1\right)_m}}
\end{eqnarray}}
The moment problem (\ref{mpn}) with the $\widetilde\rho_m$ of
(\ref{nrhopt}) can be worked out once $\epsilon_1, \dots,
\epsilon_k$ are specified. However, the degeneracy of the eigenvalue
$z=0$ of $a_{k_{\mathcal N}}$ is $q+1$.

The intrinsic nonlinear and linear CS of $H_k$ are obtained from the
corresponding ones of $H_0$ (see (\ref{cspti}-\ref{csptl})) by the
replacement $\vert\psi_m\rangle \rightarrow \vert\theta_m\rangle$.

As an illustration, a first-order SUSY transformation which
`creates' a new level at $\epsilon$ for $H_1$ is taken (for $k=q=1,
\ p=0$). The `Wronskian' is directly the solution $u(x)$ of
(\ref{ptgs}); with this input for $\mu=1.9, \ \epsilon=3/2< E_0 =
9/2$ we have drawn in Fig. 3 the SUSY partner potential (black
curve) of the P\"oschl-Teller potential with $\nu=3$ (gray curve).

\begin{figure}[ht]
\centering \epsfig{file=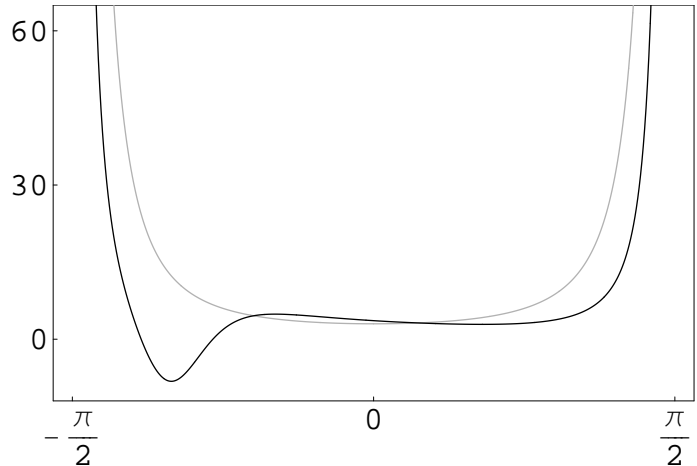, width=10cm} \caption{\small
First-order SUSY partner potential $V_1(x)$ (black curve) of the
P\"oschl-Teller potential with $\nu=3$ (gray curve) obtained by
using as seed the $u(x)$ of (\ref{ptgs}) with $\mu=1.9, \
\epsilon=3/2< E_0 = 9/2$. The new potential has an additional level
at $\epsilon$.}
\end{figure}

\section{Conclusions}

In this paper we have derived coherent states for Hamiltonians $H_k$
attained from a given initial one through the higher-order SUSY QM.
We have shown here, and previously for the harmonic oscillator
\cite{fhn94,fh99}, that it is important to determine the algebraic
structures ruling those potentials. It turns out that the intrinsic
and linear algebras of the initial Hamiltonian are inherited by its
corresponding SUSY partners in the subspace associated to the
isospectral part of the spectrum. Moreover, we have discussed an
interesting additional algebra of $H_k$ (the so-called natural)
generalizing the one which was first introduced for the SUSY
partners of the harmonic oscillator \cite{fhn94,fh99}. We have shown
as well that the natural and intrinsic algebras are deformations
from each other, and our analysis shows that the natural is more
involved that the intrinsic one. On the other hand, the linear
algebra we have studied is a deformation simplifying at maximum the
intrinsic structure of our systems. It is worth to notice that, up
to this moment, the last procedure has been elaborated at a purely
algebraic level, and it has been implemented to map somehow the
original system into the harmonic oscillator. This suggests a class
of problems which could be addressed in the future, in particular,
it would be important to analyze the consequences of this
linearization at a differential level. This is a quite interesting
problem which, as far as we know, is open.

\section*{Acknowledgments}

The authors acknowledge the support of Conacyt, projects 49253-F and
50766, and research grants from NSERC of Canada. Part of this work has
been done while VH visited the Cinvestav and DJFC the Universit\'e de
Montr\'eal. These institutions are acknowledged for hospitality and
financial support.


\end{document}